# Design of laser-electron storage ring lattice dedicated to generation of intense X-rays under Compton scattering.

P.Gladkikh
*National Science Center "Kharkov Institute of Physics and Technology", Kharkov, Ukraine*

The lattice of laser-electron storage ring with controlled momentum compaction factor dedicated to generation of intense X-rays under Compton scattering is described. In such storage ring one can achieve the large energy acceptance and keep the long-term stable motion of electron beam with large energy spread. The intensity of X-rays may be very stable due to using of electron beam with steady-state parameters for Compton scattering. Parameters of the storage ring, electron beam and X-rays obtained by electron beam dynamics simulation involving Compton scattering are presented.

## 1. Introduction

The proposal of the laser-electron storage rings (LESR) dedicated to hard X-rays generation by means of the Compton scattering of the laser light on low-energy electron beam of the storage ring was stated in 1997 [1]. For example, in order to obtain X-rays energy $\varepsilon_\gamma \approx 33$ keV for angiographic studies under Compton scattering of laser photons with energy $\varepsilon_{las} = 1.164$ eV (neodymium laser) one needs electron beam energy $E_0 \approx 43$ MeV. To generate such X-rays in synchrotron radiation source one needs electron beam energy $E_0 \approx 2.5$ GeV and superconducting wiggler with field $B \approx 7.5$ T. It is clear that X-rays generators based on Compton scattering may become inexpensive, compact sources of the intensive X-rays.

There are two basic problems in LESR design. The first one is associated with strong effect of the intrabeam scattering at low electron beam energy. Because of this effect the beam size grows quickly and Compton scattering intensity decreases. The second problem is associated with large electron beam energy spread because of fluctuation of Compton generation. The value of the energy spread may run up to few percents and one needs to keep electron beam during long term in order to achieve the high Compton beam intensity.

Two basic schemes of LESR were proposed for intensive Compton scattering. In the first scheme electron beam with non-steady-state parameters is supposed to be used. Intensive electron beam is injected from linac into storage ring and this beam is being used during short term within which beam size does not significantly increase and after that injection is repeated. For such experiments one needs linac with bunch charge $q_b \approx$ 1-2 nC, beam emittance $\varepsilon \approx$ 30-50 nm and repetition frequency $f_{inj} \approx$ 10-100 Hz. In these conditions and under developing parameters of the laser systems designers expect average scattered beam intensity $n_\gamma \approx 10^{13}$-$10^{14}$ phot/s and spectral brightness $B \approx 10^{13}$-$10^{14}$ phot/(s*mm$^2$*mrad*0.1%BW). The main imperfection of such LESR scheme is the pulse nature of radiation whereas some experiments require long-term stability of X-rays intensity (for example, biological studies, laser cooling of electron beam).

In this paper the second scheme of LESR with controlled momentum compaction factor designed at NSC KIPT is described [2]. In such storage ring one can achieve the large energy acceptance and keep the long-term stable motion of electron beam with large energy spread. The intensity of X-rays may be very stable due to using of electron beam with steady-state parameters.

* Work supported by NATO SfP/977982 grant (X-rays generator)

## 2. Main requirements for ring lattice

Under Compton scattering (CS) of laser photon on relativistic electron the energy of the scattered X-ray is determined by the following expression [3]

$$\varepsilon_{\gamma} = \frac{1+\beta\cos\varphi}{1-\beta\cos\phi}\varepsilon_{las}, \qquad (1)$$

where $\varepsilon_{\gamma}$ is the scattered quanta energy,
$\varepsilon_{las}$ is the laser photon energy,
$\varphi$ is the collision angle ($\varphi = 0$ corresponds to head-on collision),
$\phi$ is the angle between vectors of electron and X-ray velocities,
$\beta = v / c$ is the ratio of electron and light velocities.

Thus under head-on collision the quanta with maximal energy scatters towards the direction of electron velocity

$$\varepsilon_{max} = 4\,\gamma^2\,\varepsilon_{las},$$

where $\gamma$ is the Lorentz factor.

In single collision the number of scattered photons is determined by the luminosity L and the total cross-section of Thomson scattering $\sigma$

$$n_{\gamma} = L\sigma \qquad (2)$$

In laboratory frame under assumption of Gaussian distribution of densities of electron and laser beams the luminosity is described by the expression [4]

$$L = \frac{n_e n_l}{2\pi\sqrt{\left(\sigma_{ze}^2+\sigma_{zl}^2\right)}\sqrt{\left(\sigma_{xe}^2+\sigma_{xl}^2\right)+\left(\sigma_{se}^2+\sigma_{sl}^2\right)\tan^2(\varphi/2)}}, \qquad (3)$$

where $\sigma_{xe}$, $\sigma_{xl}$ are the transversal sizes of the electron and laser beams at interaction point (IP) in collision plane (in plane where the vectors of the electron and laser photon velocities are located; below we assume laser beam propagates in reference orbit plane),
$\sigma_{ze}$, $\sigma_{zl}$ and $\sigma_{se}$, $\sigma_{sl}$ are the vertical and longitudinal sizes of the electron and laser beams,
$n_e$, $n_l$ are the numbers of electrons and photons in colliding beams.

The expressions (1)-(3) define completely the requirements for electron and laser beam parameters which determine the energy range and intensity of the scattered X-rays.

Let us carry out the estimations of the Compton beam intensity from compact storage ring with circumference C = 15 m (revolution frequency is equal to 20 MHz) for collision angle $\varphi = 10°$ and for modern electron and laser beam parameters. The most perspective laser for using in LESR is the neodymium laser which generates photons with energy $\varepsilon_{las} = 1.164$ (2.328) eV. The modern lasers operate at repetition frequency $f_{rep} \approx 350$ MHz, pulse duration $\tau_p \approx 10$ ps and average power $P \approx 10$ W ($\approx 30$ nJ or $1.8 \cdot 10^{12}$ photons are generated during single laser pulse). We assume the transversal laser beam size at IP $\sigma_{xl} = \sigma_{zl} = 50\ \mu$, longitudinal one $\sigma_{sl} = 1.5$ mm and we also assume electron and laser beam sizes are coinciding. Under these parameters in single collision we obtain the probability of the Compton scattering equal approximately to $8 \cdot 10^{-11}$ per electron. The electron bunch containing 2 nC ($\approx 40$ mA per bunch of the stored current) generates $n_{\gamma} \approx 2 \cdot 10^7$ phot /s. It is not enough for most of researches. To increase the scattered beam intensity we need to increase the number of the electron bunches and laser photons number. We can essentially increase the laser flash energy by use of the optical cavity for laser pulses stacking. The obtaining of the colliding beams with sizes less than the ones above mentioned is a very complicated task. Note, that longitudinal size essentially decreases the Compton beam intensity in a case of non-head-on collision, because the transversal size is as a rule much less than the longitudinal one. As is obvious from expression (3) the criterion of the collision angle smallness is determined by the expression

$$\tan^2(\varphi / 2) < \sigma_x / \sigma_s \tag{4}$$

The less is the longitudinal size the higher is the permissible value of the collision angle without essential decreasing of the scattering intensity.

As appears from above presented estimation the requirements for storage ring lattice suitable for obtaining of the intensive scattered beam are following:

- the transversal size of the electron beam at IP should be less than several tens of microns;
- the longitudinal one should be less than several millimeters;
- the bunch charge should be greater than 1 nC;
- the structure of the interaction region should allow the collision angle $\varphi \to 0$.

In practice it is very difficult to fulfil those requirements because of the details of the electron beam dynamics.

## *3. Beam dynamics in LESR*

### 3.1. Details of beam dynamics in LESR.

The main designing problems of the LESR dedicated for generation of intensive X-rays in steady-state operation mode are associated with large steady-state energy spread of the electron beam. As a result of the CS an electron loses significant part of energy what causes strong excitation of the synchrotron oscillations because of stochasticity of the scattering. The partial energy spread due to CS is described by expression [4]

$$\delta_{CS} = \sqrt{\frac{2}{3}\gamma\frac{\varepsilon_l}{\varepsilon_0}}\,, \tag{5}$$

where $\varepsilon_0$ is the rest electron energy. For example, at $\gamma = 100$ ($E_0 \approx 50$ MeV) $\delta_{CS} \approx 1.23$ %. Due to combined effect of the synchrotron radiation (SR) and CS the total steady-state energy spread is

$$\delta_{tot} = \left[\left(\frac{\tau_{tot}}{\tau_{CS}}\right)\delta_{CS}^{\;2} + \left(\frac{\tau_{tot}}{\tau_{SR}}\right)\delta_{SR}^{\;2}\right]^{1/2}, \tag{6}$$

where $\tau_{tot} = 1/(\tau_{CS}^{-1}+\tau_{SR}^{-1})$ is the total damping time of the synchrotron oscillations,
$\tau_{CS} \approx E_0T_{rev} /\Delta E_{CS}$, $\tau_{SR} \approx E_0T_{rev} /\Delta E_{SR}$ are the partial damping times because of energy losses due to CS and SR, correspondingly,
$\Delta E_{CS}$, $\Delta E_{SR}$ are the average energy losses per turn because of CS and SR, correspondingly,
$\delta_{SR}$ is the partial energy spread caused by SR,
$T_{rev}$ is the revolution time.

Under intensive CS when energy losses caused by CS are comparable to those ones because of SR the steady-state energy spread value reaches a few percents. To obtain the stable electron beam motion we need to solve several serious problems.

*In first*, to obtain the acceptable quantum life time we need the large energy acceptance of the storage ring and unreasonable RF-voltage may be required. There is no place in compact storage ring for the placement of the great number of the RF-cavities.

*In second*, the transversal and longitudinal beam dynamics are determined in this case not only by linear on momentum deviation effects but also by quadratic and higher order ones. The aberrations do not allow to focus the electron beam at IP what causes the decreasing of the CS intensity. Besides, the strong chromatic effects cause the beam diffusion because of synchrobetatron resonance at high RF-voltage. The quadratic on momentum deviation terms become apparent in longitudinal dynamics as the distortion of the separatrix shape and reducing of the RF-acceptance. Thus, we need the possibilities of the suppression of aberrations in ring lattice.

*In third*, the effects of the intrabeam scattering (IBS) become very strong at low electron beam energies. The emittances growth comparably to natural emittances may reach 2-3 orders what causes the essential increasing of the beam size and significant decreasing of the CS intensity.

And finally, *in fourth*. The sextupole lenses are used in storage rings to correct the chromatic effects. The natural chromaticity of the storage ring is very large under condition of the strong focusing of the electron beam at IP (compact storage ring with low-β insertion) and the required sextupole strengths also become large. Dynamics aperture of the ring (DA) is reduced and the problem of the obtaining of the acceptable DA should also be solved at lattice design.

### 3.2. The transversal beam dynamics

The displacement of the electron orbit from reference one is determined by dispersion functions of the storage ring

$$\Delta x = \eta_1\delta + \eta_2\delta^2 + \ldots \tag{7}$$

where $\eta_1$ and $\eta_2$ are the linear and second order dispersion functions, correspondingly [5]

$$\eta_1(s) = \frac{1}{2\sin\pi Q_x}\int_s^{s+C}\frac{\sqrt{\beta_x(s)\beta_x(\sigma)}}{\rho(\sigma)}\cos(\pi Q_x - \mu_{\sigma s})d\sigma \tag{8}$$

$$\eta_2(s) = -\eta_1(s) + \frac{1}{2\sin\pi Q_x} * \int_s^{s+C}\sqrt{\beta_x(s)\beta_x(\sigma)}\cos(\pi Q_x - \mu_{\sigma s}) * \left(K_1(\sigma)\eta_1(\sigma) - \frac{1}{2}K_2(\sigma)\eta_1^2(\sigma)\right)d\sigma \tag{9}$$

In those expressions $K_1$ and $K_2$ are the quadrupole and sextupole strengths, $Q_x$ and $\mu_x$ are the horizontal betatron frequency and phase advance. The dispersion causes the growth of the effective beam size so both dispersions must be suppressed at interaction point $\eta_1 = \eta_2 = 0$. As we have mentioned above both quadrupole and sextupole strengths in LESR are large so the second order dispersion may be large too as at IP so all over the ring. The suppression of the first order dispersion $\eta_1 = 0$ is only performed by using the linear lattice elements (bending and quadrupoles). To suppress the second order dispersion we also need to phase the sextupoles placed over azimuth with the non-zero first order dispersion, as appears from (9).

The betatron frequencies and amplitude functions also depend on momentum deviation because of changing of the strengths of lattice elements. Those dependences are characterized by derivatives $\partial Q_{x,z} / \partial\delta$, $\partial^2 Q_{x,z} / \partial\delta^2$, …, $\partial\beta_{x,z} / \partial\delta$, …The changes of the betatron amplitudes and frequencies cause the nonlinearity of the betatron motion. In this connection the effective emittance increases. Besides, the nonlinear dependence of the betatron tuning on momentum deviation

$$\Delta Q_y = \frac{\partial Q_y}{\partial\delta}\delta + \frac{\partial^2 Q_y}{\partial\delta^2}\delta^2 + \frac{\partial^3 Q_y}{\partial\delta^3}\delta^3 + \ldots \tag{10}$$

may stimulate betatron resonances even if natural chromaticity is compensated $\partial Q_y / \partial\delta = 0$.

That is why we need the possibility of the correction of nonlinear terms (10) in ring lattice.

### 3.3. Longitudinal electron beam dynamics

The most appropriate scheme of LESR is the raystrack with long straight sections for the placement of the injection system, RF-cavity and optical cavity (Fig.1)

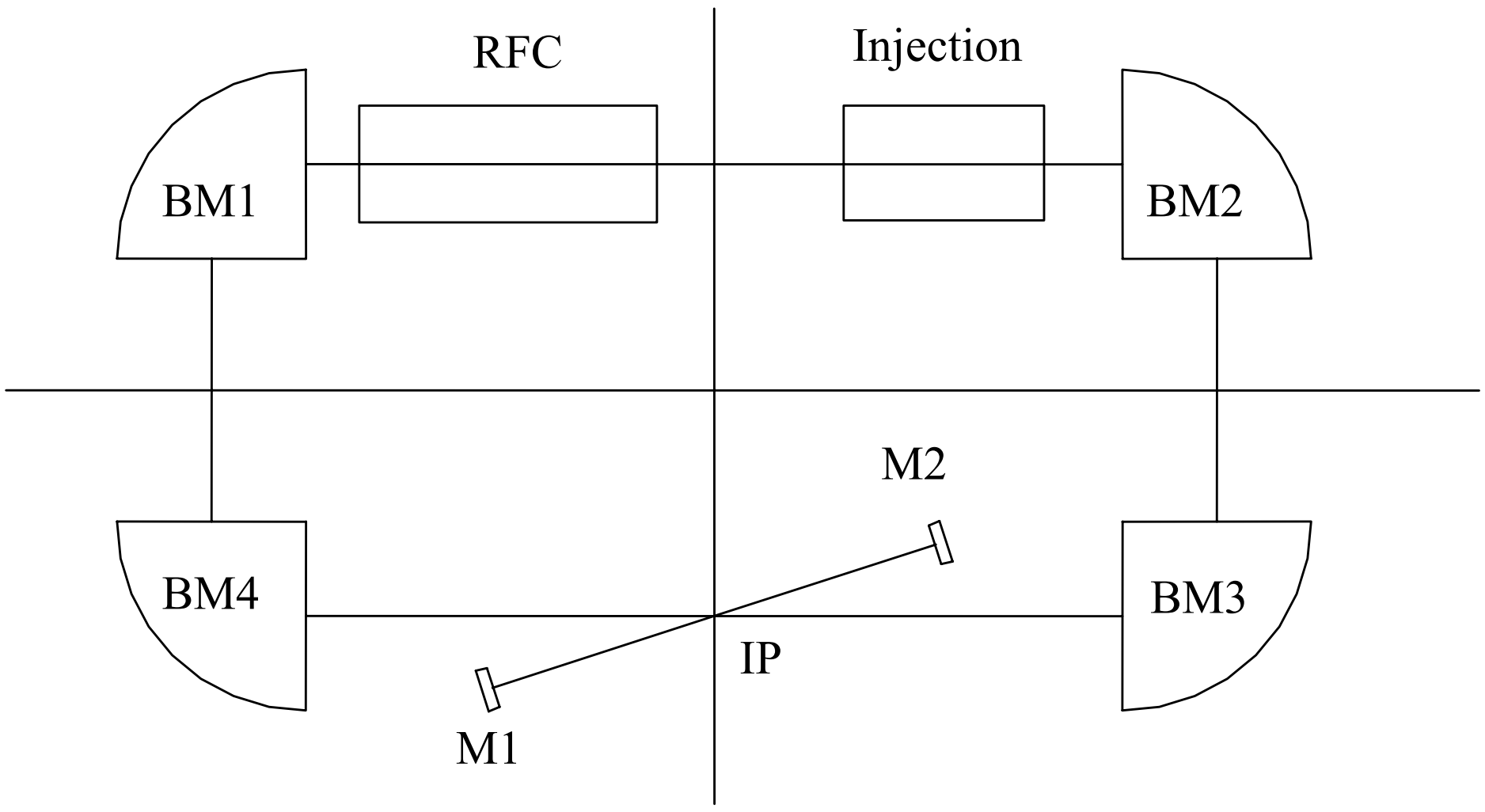

Fig.1. Compact LESR scheme. BM1-BM4 are the bending magnets, IP is the interaction point, M1-M2 are the optical mirrors, RFC is the RF-cavity.

The long straight section with IP must be dispersion free, otherwise the CS intensity will be low. The section with the RF-cavity must be also dispersion free, otherwise synchrobetatron resonances will be excited and electron beam will be lost. The bending angle of the compact storage ring must be large $\varphi_{BM} \gtrsim \pi/2$ and as a result the dispersion on bending exit will be approximately equal to bending radius $\eta_1 \approx \rho_{BM}*(1-\cos\varphi_{BM}) \approx \rho_{BM}$ (for bendings with zero field index this expression is just correct). Consequently, the first order momentum compaction factor $\alpha_1$ will be large

$$\alpha_1 = \frac{1}{C}\int_{BM} \frac{\eta_1}{\rho_{BM}} ds \qquad (11)$$

For ring with $C \approx 15$ m, bending radius $\rho_{BM} = 0.5$ m and bending angle $\varphi_{BM} = 90°$ momentum compaction factor is about of 0.1. The RF-acceptance at large RF-voltage is inversely proportional to square root from momentum compaction factor

$$\sigma_{RF} \approx \sqrt{\frac{2eV_{RF}}{\pi\alpha_1 h E_0}} \qquad (12)$$

where $V_{RF}$ is the RF-voltage, h is the harmonics number. For example, in order to obtain acceptance $\sigma_{RF} = 5$ % we need $V_{RF} \approx 1.18$ MV at electron beam energy $E_0 = 100$ MeV, harmonics number h = 30 and momentum compaction factor $\alpha_1 = 0.1$. Thus, it is very important to minimize the momentum compaction factor at lattice design.

Longitudinal electron beam size determining the CS intensity under non-head-on collision also depends on momentum compaction factor

$$\sigma_s = \sqrt{\frac{\alpha_1 h}{2\pi|\cos\phi_s|}\frac{E_0}{eV_{RF}}}\lambda_{RF}\delta_{tot} \qquad (13)$$

Here $\phi_s$ is the synchronous phase, $\lambda_{RF}$ is the RF-wavelength. For example, to obtain the longitudinal beam size $\sigma_s = 10$ mm under energy spread $\delta = 0.5$ % and above described parameters of the storage ring and electron beam we need RF-voltage $V_{RF} \approx 2.1$ MV. It is practically impossible to provide such voltage in compact storage ring, thus it is very important to reach the momentum compaction factor as small as possible.

In case of both large momentum deviation and second order dispersion the quadratic term of the transversal displacement (7) becomes comparable to linear one. This effect causes quadratic terms in orbit lengthening, in other words the momentum compaction factor becomes dependent on momentum deviation [6, 7]

$$\alpha = \frac{d\frac{\Delta C}{C}}{d\frac{\Delta p}{p}} = \alpha_1 + \alpha_2\delta + ... \tag{14}$$

where $\alpha_2$ is the second order momentum compaction factor.

The simple geometric consideration shows that orbit lengthening

$$\frac{\Delta C}{C} = \frac{1}{C}\oint\left(\frac{1+\Delta x/\rho}{\cos(x')} - 1\right)ds = \alpha_1\delta + \alpha_2\delta^2 + ... \tag{15}$$

Substituting (7) in (15) we obtain the following expressions

$$\alpha_1 = \frac{1}{C}\int_0^C \frac{\eta_1}{\rho}ds, \quad \alpha_2 = \frac{1}{C}\int_0^C\left(\frac{\eta_2}{\rho} + \frac{\left(\eta_1'\right)^2}{2}\right)ds \tag{16}$$

Longitudinal motion equations taking into account quadratic terms of the orbit lengthening and neglecting damping of the oscillations are described as

$$\dot{\phi} = \omega_{RF}\alpha = \omega_{RF}\left(\alpha_1\delta + \alpha_2\delta^2\right),$$

$$\dot{\delta} = \frac{eV_{RF}}{E_0 T_{rev}}\left[\sin(\phi_s + \phi) - \sin\phi_s\right], \tag{17}$$

One can obtain these equations from Hamiltonian

$$H = \omega_{RF}\left(\frac{\alpha_1}{2}\delta^2 + \frac{\alpha_2}{3}\delta^3\right) + \frac{eV_{RF}}{E_0 T_{rev}}\left[\cos(\phi_s + \phi) + (\phi_s + \phi)\sin\phi_s\right] \tag{18}$$

The fixed points are determined by system of equations

$$\frac{\partial H}{\partial \phi} = 0, \frac{\partial H}{\partial \delta} = 0 \tag{19}$$

There are one stable fixed point $\phi = \phi_s$, $\delta = 0$ in linear case $\alpha_2 = 0$ and one unstable fixed point $\phi = \pi - \phi_s$, $\delta = 0$. The particles oscillate inside separatrix around stable fixed point. Quadratic on momentum deviation term causes additional stable $\phi = \pi - \phi_s$, $\delta = -\alpha_1 / \alpha_2$ and unstable fixed points. The beam oscillating inside separatrix around $\phi_s$ is called as normal beam, the beam oscillating around $\pi - \phi_s$ is called as anomalous one. The stable phase of the anomalous beam is located on the RF-wave flank where the normal beam is unstable. Generally, oscillations are stable if $d\phi / dt$ changes sign periodically (the first equation 17). In case of the normal beam $\alpha_2\delta^2 << \alpha_1\delta$ and derivative sign is changed because $\delta$ oscillates around $\delta = 0$. In anomalous case one can see from (14) and first equation (17) that $d\phi / dt$ may change sign if $\delta$ oscillates around equilibrium value $\delta = - \alpha_1 / \alpha_2$. The nature of the phase trajectories is strongly dependent on relation of linear and quadratic momentum compaction factors. Quantitatively those relations are described by the value of the critical quadratic momentum compaction factor

$$\alpha_{2C} = \sqrt{\frac{E_0\omega_{RF}T_{rev}|\alpha_1|^3}{12eV_{RF}\left[-\cos\phi_s + \left(\frac{\pi}{2} - \phi_s\right)\sin\phi_s\right]}} \tag{20}$$

When $|\alpha_2| << \alpha_{2C}$ the phase trajectories of the normal beam are similar to those ones in linear theory. The trajectories distort and the asymmetry of the separatrix brunches arise when $|\alpha_2|$ increases. In the case of $|\alpha_2| > \alpha_{2C}$ the separatrixes of the normal and anomalous beams become $\alpha$ - like (Fig.2). The energy acceptance is described as following

$$\sigma_{RF} = \frac{3}{2}\left|\frac{\alpha_1}{\alpha_2}\right| \tag{21}$$

and $\sigma_{RF}$ decreases quickly under $|\alpha_2|$ increasing.

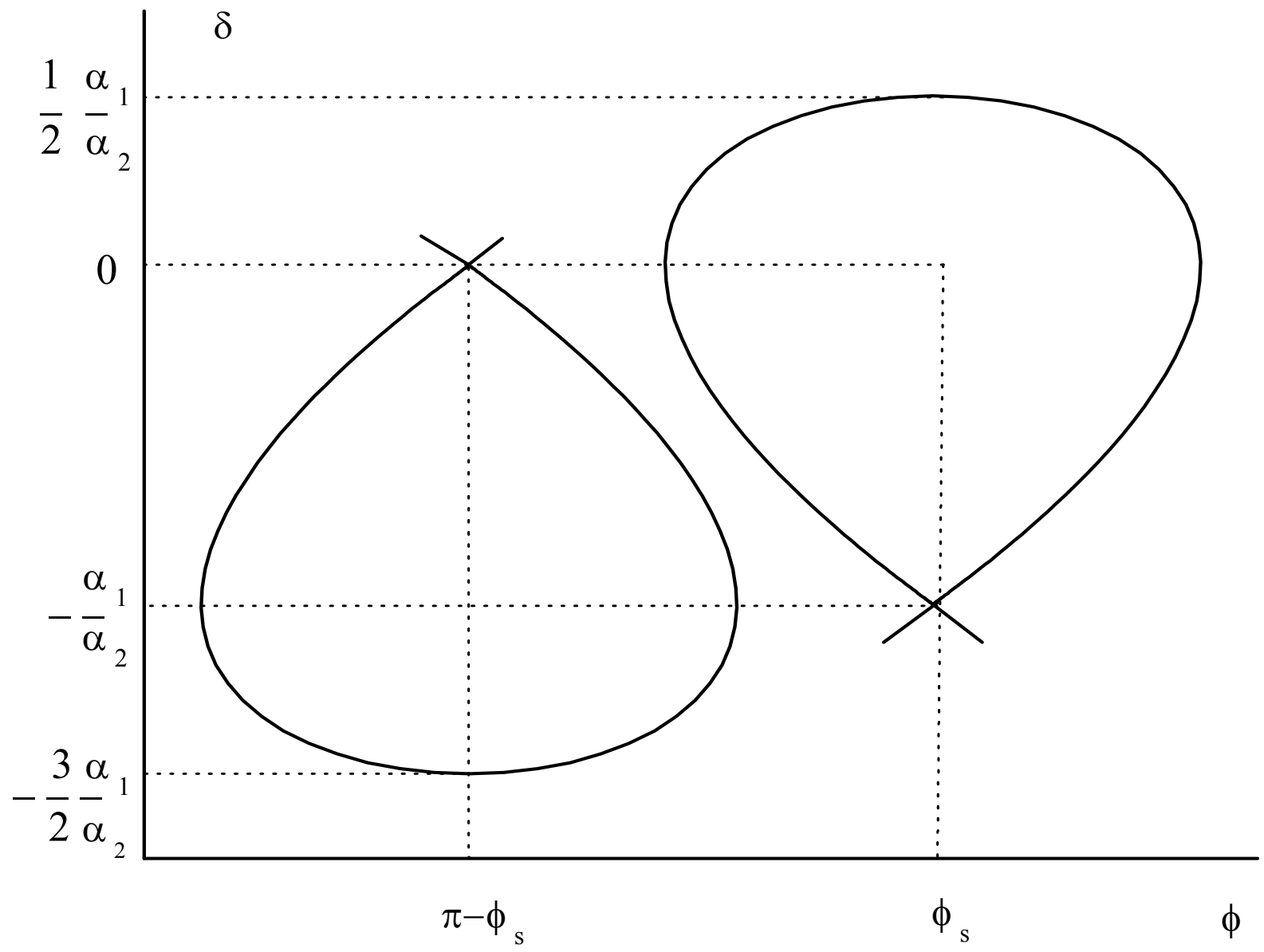

Fig.2. Separatrixes of the normal and anomalous beams in case $|\alpha_2| > \alpha_{2C}$.

Thus, we need to control the second order dispersion over beam orbit in storage ring dedicated to keep the electron beam with large energy spread. Otherwise, it is impossible to provide the large RF-acceptance under operation mode with low momentum compaction factor.

### 3.4. Intrabeam scattering (IBS)

In frame bound with electron beam the particles oscillate in transversal plane and they may scatter one another. The longitudinal component of the particle pulse arises as a result of such collision. When this component is large the colliding particles may abort from separatrix and may be lost (Touchek effect). The stochastic collisions with small transmitted pulses cause the emittances growth with corresponding times [8].

$$\tau_x^{gr} = \tau_x^{gr}(n_b, \varepsilon_x, \varepsilon_z, \varepsilon_s, \gamma),\ \tau_z^{gr} = \tau_z^{gr}(n_b, \varepsilon_x, \varepsilon_z, \varepsilon_s, \gamma),\ \tau_s^{gr} = \tau_s^{gr}(n_b, \varepsilon_x, \varepsilon_z, \varepsilon_s, \gamma).$$

The steady-state transversal and longitudinal emittances are formed as a result of combined effect of the IBS and radiation cooling. As a rule the vertical emittance is determined by coupling of oscillations. The growth rates at low electron beam energies are very large comparably to natural emittances, about of dozens and hundreds times. For example, in Fig.3 the dependences of the transversal and longitudinal emittances on electron beam energy in storage ring NESTOR designed at NSC KIPT are presented.

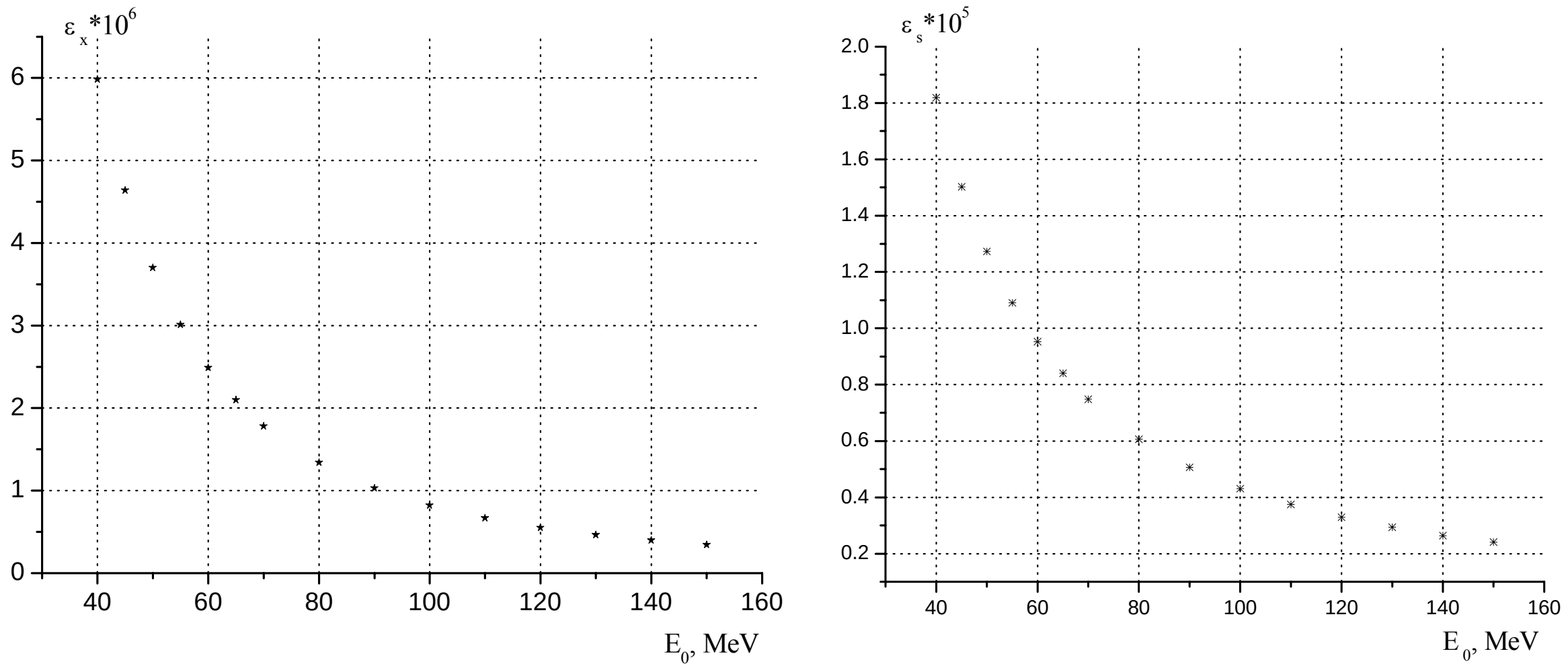

Fig.3. Dependencies of horizontal and longitudinal emittances on electron beam energy at stored bunch current $I_{stor}$ = 10 mA.

It is a very complicated analytical task to take into account the intrabeam scattering in beam dynamics involving the Compton scattering because it is the consistent problem. We simulate the IBS by using the following algorithm:

- the growth rates $\tau_y^{gr}$, y = {x, z, δ) are computed before simulation over estimated range of the beam emittances;
- the simulating element is incorporated in ring lattice and angular coordinates and momentum deviation of the particle are changed in this element by using of matrix transformations

$$y_{fin}' = y_{ini}'*(1 + T_{rev} / \tau_y^{gr}), \quad y' = \{x', z', \delta) \qquad (22)$$

During simulation the growth rates are corrected in accordance with the beam emittances. If correction time is much less than growth rate we get the beam emittances coinciding well with analytical estimations. The results of the IBS simulation in storage ring NESTOR are presented in Fig.4.

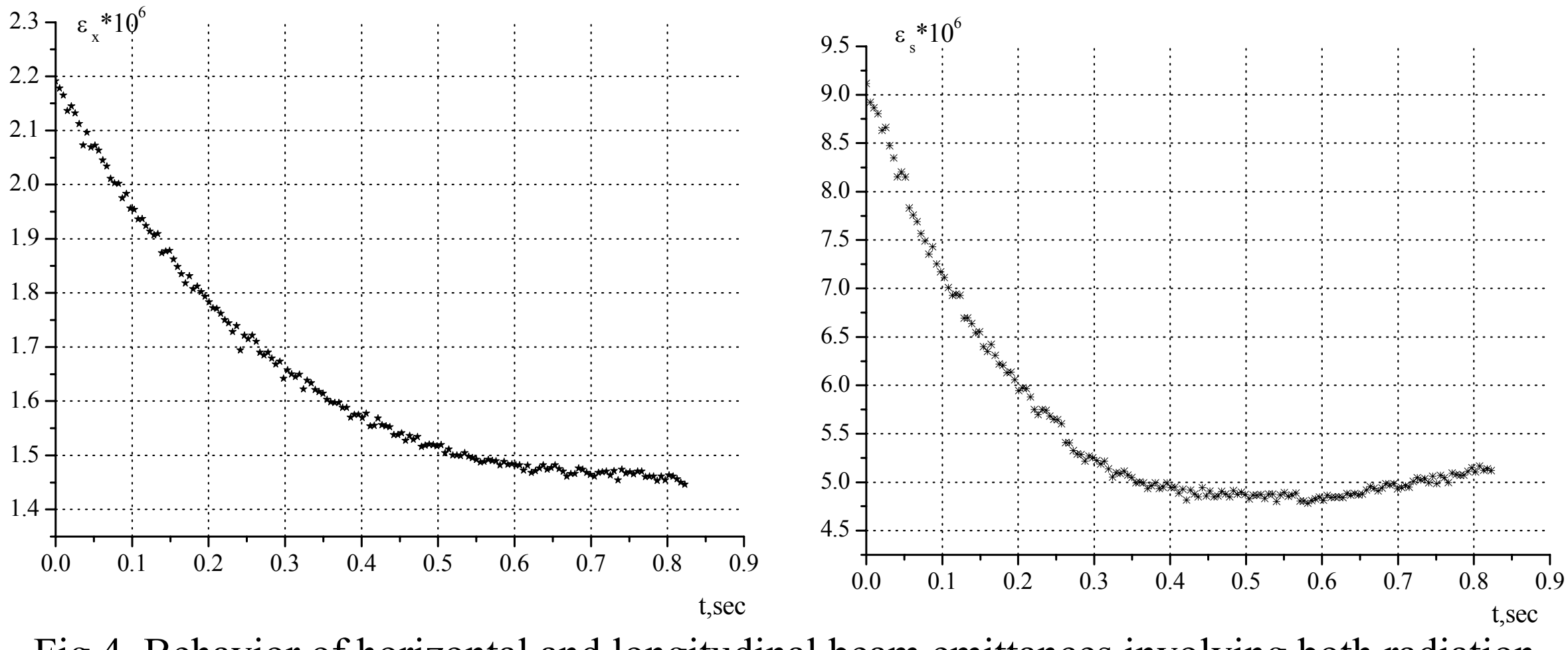

Fig.4. Behavior of horizontal and longitudinal beam emittances involving both radiation damping and IBS.

The simulation parameters were following:

- electron beam energy $E_0$ = 75 MeV;
- particles number N = 400;
- stored bunch current $I_{stor}$ = 10 mA;
- coupling coefficient $\kappa = 0.05$.

The coupling of the transversal oscillations is simulated by rotated quadrupole lens. The horizontal and longitudinal damping times at simulation energy are equal to 1.36 s and 0.7 s, accordingly, analytical estimations of the steady-state emittances are equal to $\varepsilon_x = 1.54*10^{-6}$ and $\varepsilon_s = 6.69*10^{-6}$.

## 4. Lattice of the laser-electron storage ring NESTOR. Operation modes, electron beam and X-rays parameters

Taking into account above stated we can formulate the main requirements for lattice of the ring dedicated to generation of the intense beam of the Compton scattered photons:

- amplitude functions at interaction point must be as small as possible;
- momentum compaction factor must be as small as possible, too;
- number of sextupoles in ring lattice must be enough to correct the chromatic effects and dynamics aperture.

From Fig.1 one can see, that we can decrease the momentum compaction factor if we get the negative dispersion function on beam orbit in bending magnet by using guadrupole lens between bendings. In this case one long straight section with IP will be dispersion free, dispersion function on opposite straight section will be non-zero. RF-cavity is placed on IP-drift, the injection system is placed on opposite drift. It is expediently to use the quadrupole quadruplet with final triplet to focus the electron beam at IP. Such lattice allows us to get the minimal both horizontal and vertical amplitude functions at IP and to obtain the required drift length for RF and injection systems.

It is expediently to use the separated quadrupole lens in ring arc. It allows to guarantee the appropriate conditions for chromaticity correction and to keep ring compactness. The bendings should be with non-zero field index in order to obtain the vertical focusing on ring arc.

Taking into account all above described we designe the lattice of storage ring NESTOR. Its layout is presented in Fig.5.

Bending radius and bending angle are equal to $\rho = 0.5$ m, $\varphi_{BM} = 90°$, correspondingly, field index is equal to $n = 0.6$. The maximal magnetic induction is equal to $B_{max} = 1.5$ T at the maximal electron beam energy $E_{0max}$ = 225 MeV. The length of quadrupoles is equal to 150 mm, maximal quadrupole gradient is equal to $G_{max} \approx 25$ T/m. The strong sextupoles effect on beam dynamics similarly octupoles. To correct such effect four combined sextupole lenses with octupole fields are incorporated in ring lattice. The length of all sextupole and combined lenses is equal to 100 mm. The maximal length of the drift spaces is approximately equal to 1.2 m what allows to place the 700 MHz RF-cavity and injection system elements on those drifts. Ring circumference is equal to C = 15.418 m, harmonics number is h = 36 and one can store 1, 2, 3, 4, 6, 9, 12, 18 and 36 bunches on beam orbit.

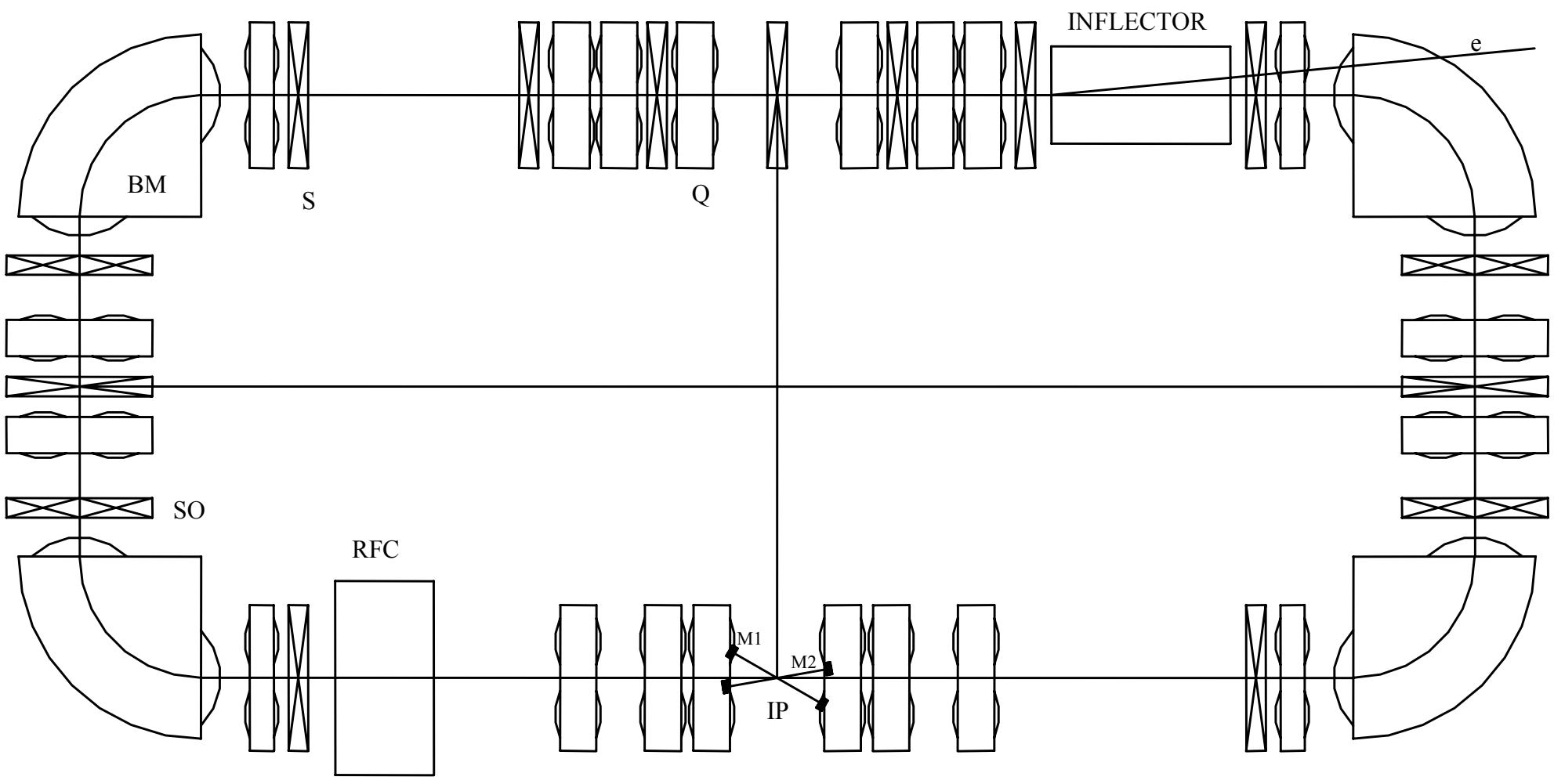

Fig.5. NESTOR layout. IP is the interaction point, BM are the bending magnets, Q, S, SO are the quadrupole, sextupole and combined sextupole and octupole lenses, correspondingly, RFC is the RF-cavity, Inflector is the injection kicker magnet, M1-M2 are the mirrors of the optical cavity.

The amplitude functions at half of ring lattice are presented in Fig.6 (ring is dissymmetrical relatively interaction point, curves begin from IP). Different focusing on IP-drift and opposite one (because dispersion functions on these drifts are different) causes insignificant asymmetry of the amplitude functions on arc and long straight sections. The less is the β-functions asymmetry the less are the amplitudes of the azimuthal perturbation harmonics and the more is the dynamics aperture of the ring.

The first order dispersion function $\eta_1$ at half of ring lattice is presented in Fig.7. We can obtain either both dispersion free long straight sections or one of them by controlling the separated quadrupole strength on ring arc. In first case we obtain the operation mode with large momentum compaction factor $\alpha_1 = 0.078$ (BM-mode). In second case dispersion function $\eta_1$ is negative on orbit in one of the arc bendings and we obtain operation mode with decreased momentum compaction factor (LM-mode). Dispersion function in Fig.7 corresponds to $\alpha_1 = 0.019$. The maximal value of the dispersion function on arc is equal to $\eta_{1max} \approx 1.2$ m, its maximal value on long straight section is equal to 0.3 m. The lattice is very flexible and it allows us to change the momentum compaction factor over wide range without betatron detuning and without essential change of the amplitude functions (we can decrease $\alpha_1$ down to zero and make it negative).

In LM-mode the first order dispersion is equal to zero only at IP-drift (approximately on one third of ring circumference), all other ring sections are not dispersion free. It allows us to place the required number of sextupoles at dispersion sections in order to suppress the second order dispersion $\eta_2$ at IP and to minimize it at all ring. It is impossible to solve this problem in BM-mode. This statement is illustrated in Fig.8 where the trajectories of the particle with large momentum deviation in both operation modes are presented.

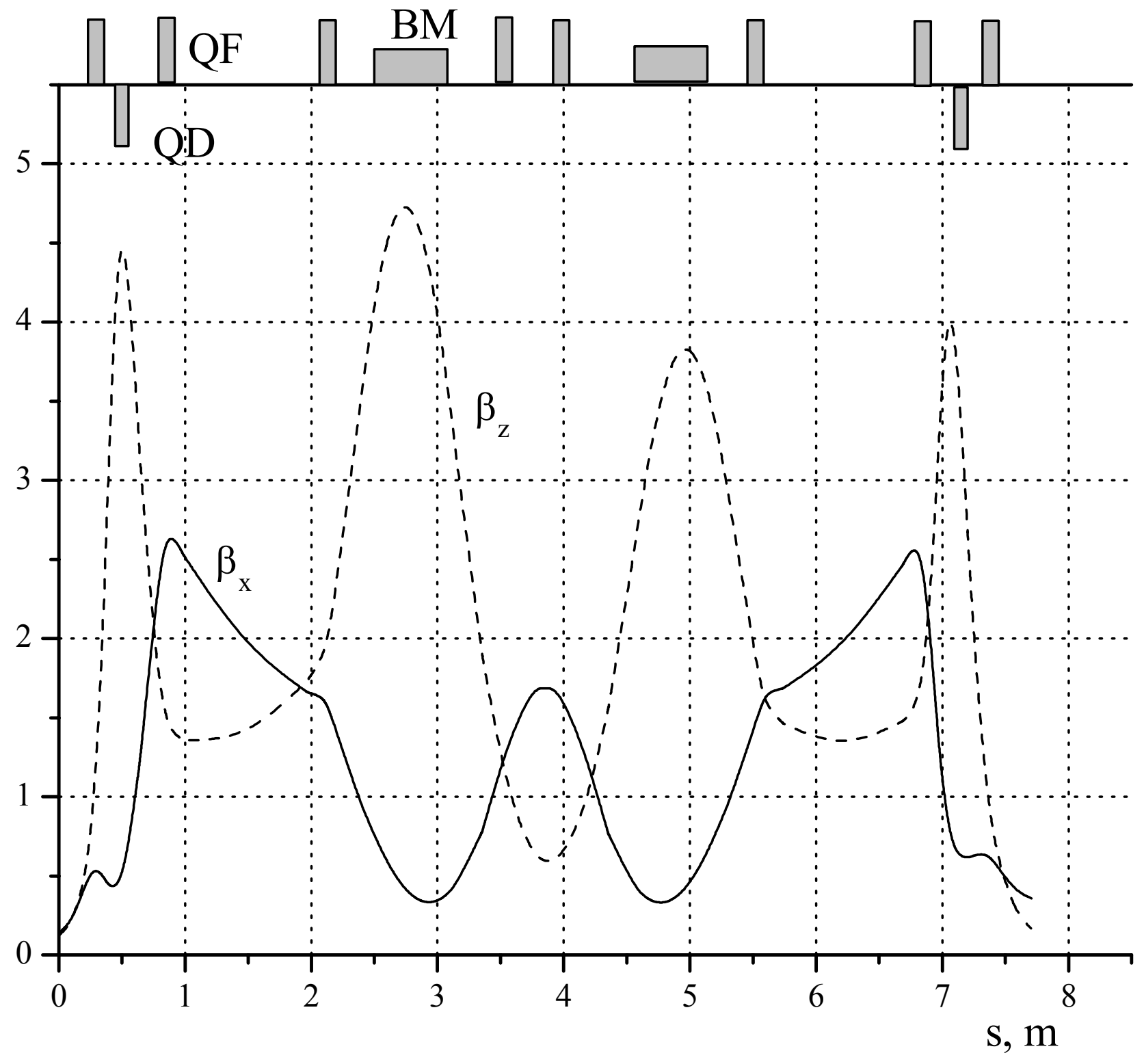

Fig.6. Horizontal $\beta_x$ and vertical $\beta_z$ amplitude functions at half ring lattice.

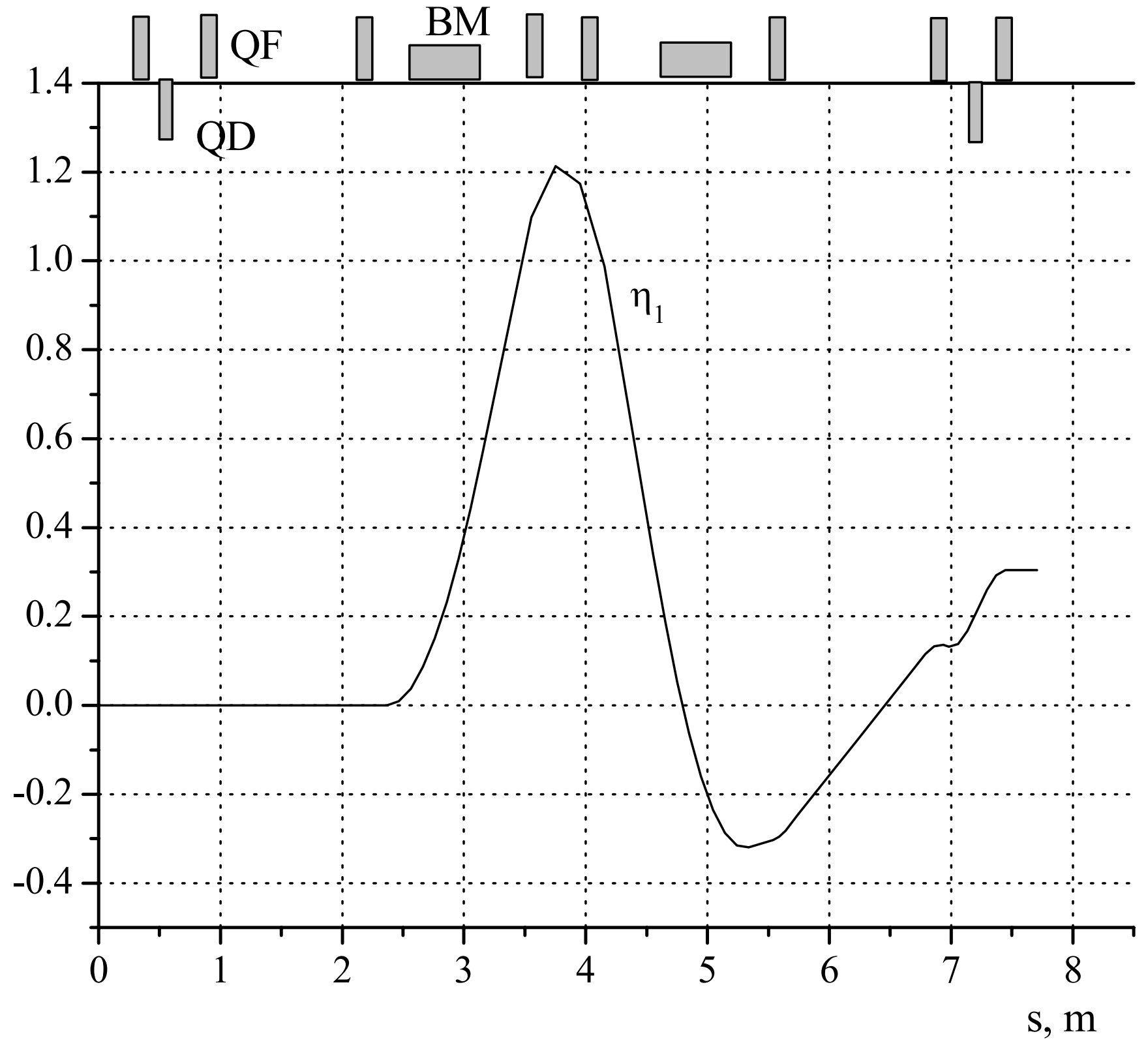

Fig.7. First order dispersion function $\eta_1$ at half of storage ring lattice

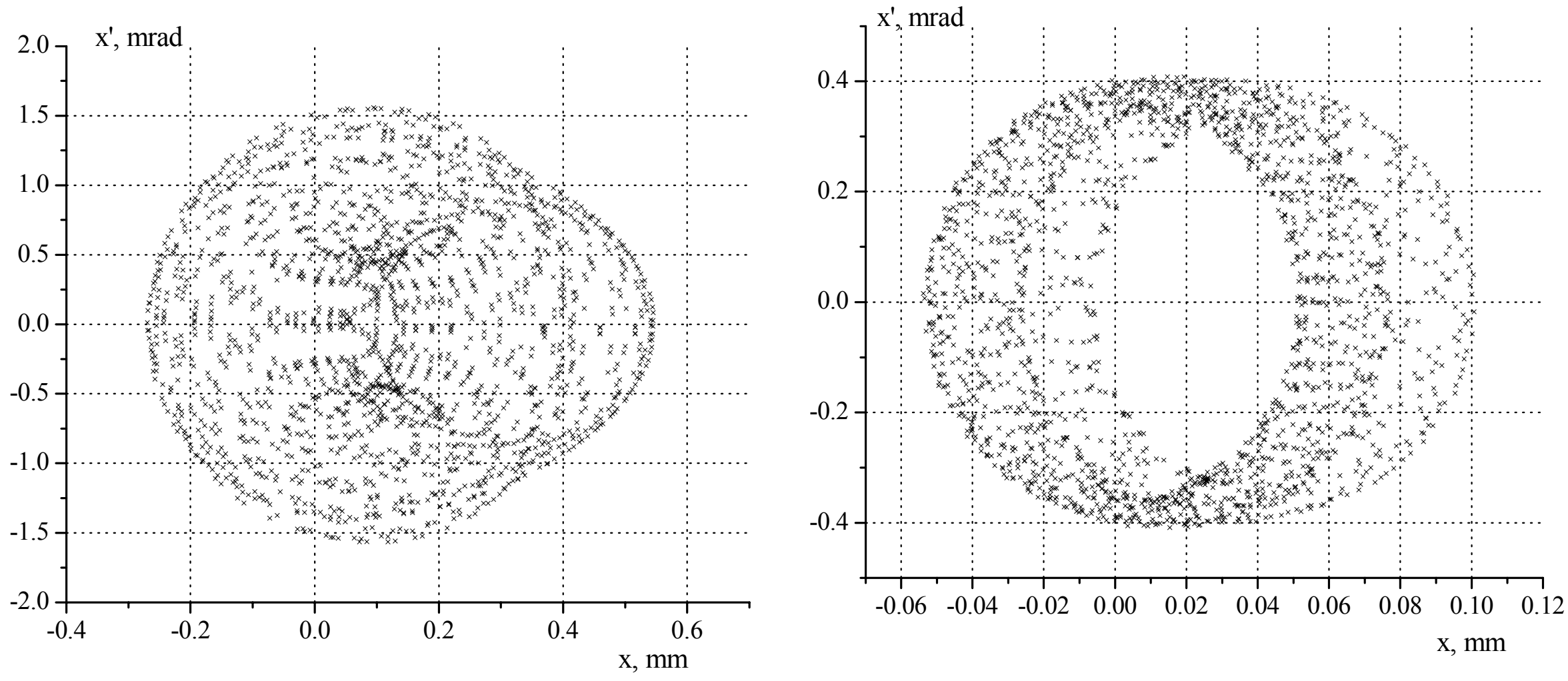

Fig.8. Horizontal phase trajectories of particle with large momentum deviation at IP azimuth in operation modes with large and low momentum compaction factor ($\alpha_1 = 0.078$ and $\alpha_1 = 0.019$, correspondingly). Initial particle coordinates are
$x_{ini} = z_{ini} = 0.1$ mm, $x'_{ini} = z'_{ini} = s_{ini} = 0$, $\delta_{ini} = 0.01$

In this figure one can see that particle trajectory in BM-mode depends essentially on momentum deviation and this effect causes growth of the effective emittance. Besides, in this mode electron beam may be slowly exited and may be lost on synchrobetetron resonances if the RF-cavity is placed at azimuth with non-zero dispersion. We observed this phenomenon in simulations.

Particle trajectory in LM-mode depends insignificantly on momentum deviation. The quadratic dispersion at IP is practically suppressed (Fig.9) and the value of the second order momentum compaction factor is small, $\alpha_2 = 0.38$. Its critical value at electron beam energy $E_0 = 225$ MeV and RF-voltage $V_{RF} = 0.3$ MV is $\alpha_{2C} = 0.33$, consequently $|\alpha_2| > \alpha_{2C}$. The separatrix of the longitudinal motion in these conditions is presented in Fig.10. One can see in this figure that separatrix shape is distorted because for even such small second order momentum compaction factor the quadratic on momentum deviation terms strongly disturb electron beam dynamics. Nevertheless, the value of the RF-acceptance is more than 7% at maximal electron beam energy for such relations between linear and quadratic momentum compaction factors.

The electron beam motion is really stable over the momentum deviation range, which is determined in chosen lattice by nonlinear shift of the betatron frequencies of the particles with large amplitudes. Dependences of the betaron frequencies on momentum deviation are presented in Fig.11. We have obtained these dependences from beam dynamics simulation. In this figure one can see that betarton detuning becomes dangerous for momentum deviation $\delta > 3$ % because particles may cross the great number of the resonances. The dynamics simulation of such particles shows that betatron amplitudes increase and particles are lost. Nevertheless, such RF-acceptance is sufficient for obtaining of the scattered beam intensity $n_\gamma \approx 10^{15}$ / s over all electron beam energy range.

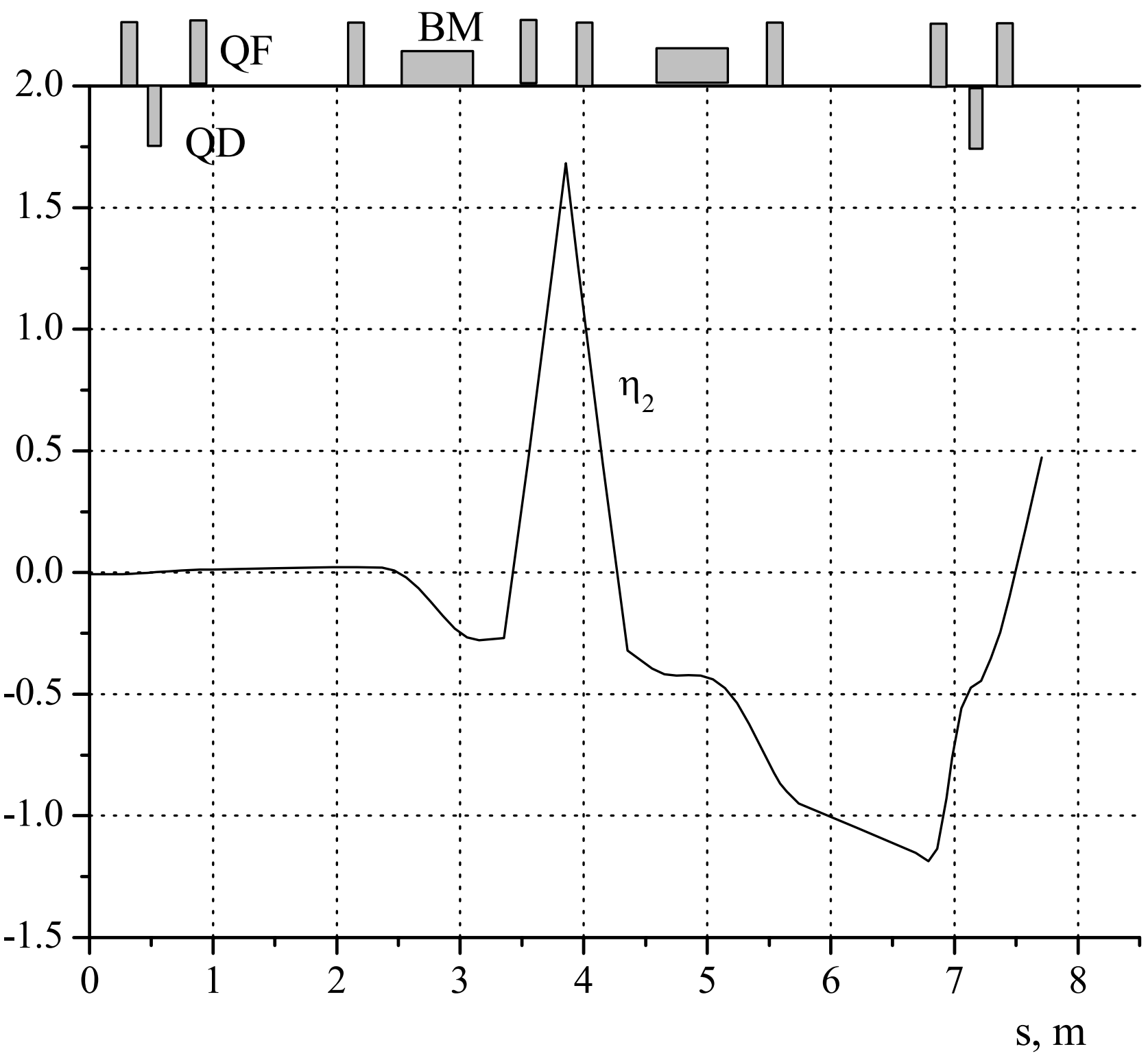

Fig.9. The second order dispersion at half of ring lattice.

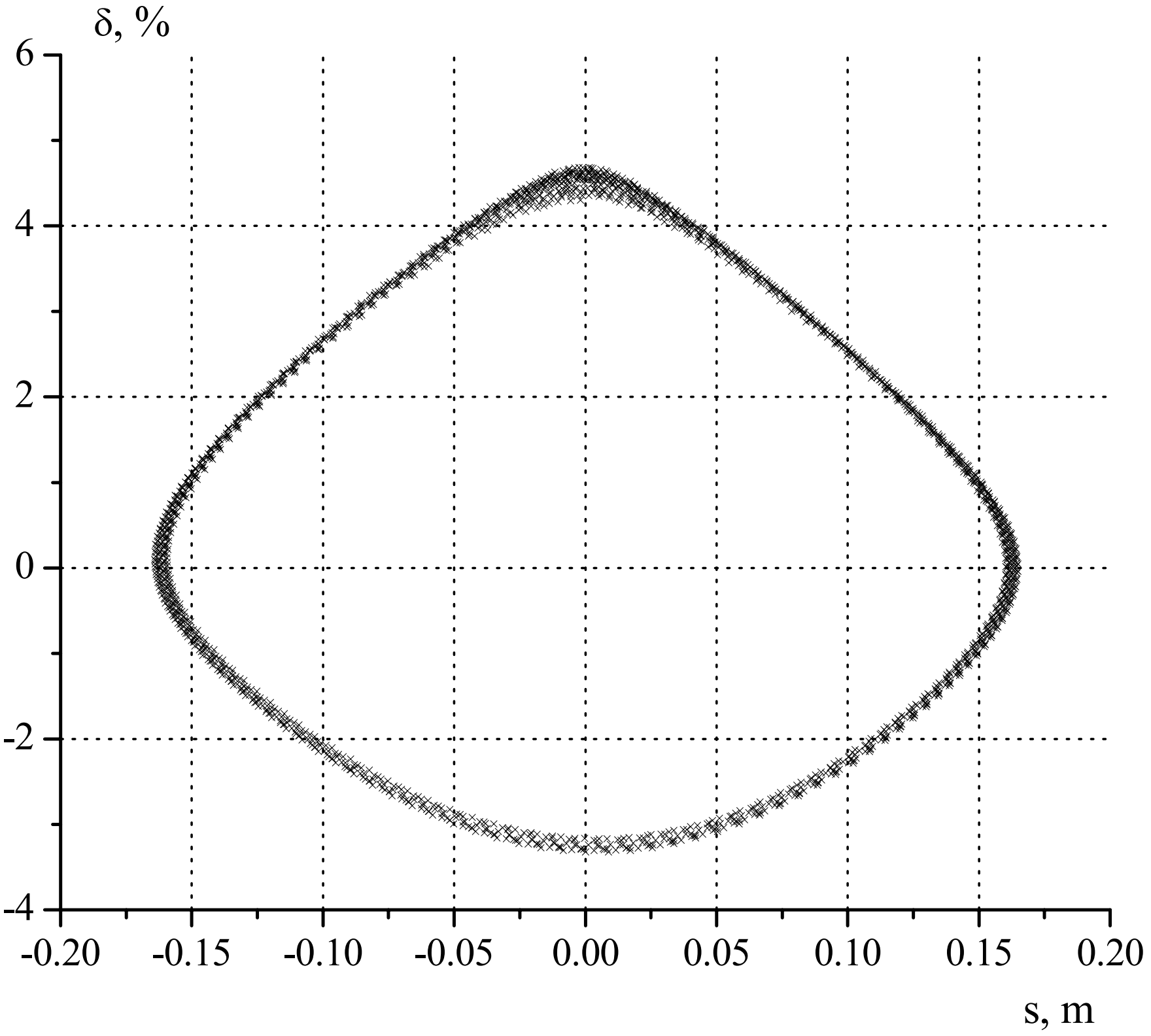

Fig.10. Separatrix of longitudinal motion at electron beam energy $E_0$ = 225 MeV and RF-voltage $V_{RF}$ = 0.3 MV. Momentum compaction factor $\alpha_1$ = 0.019.

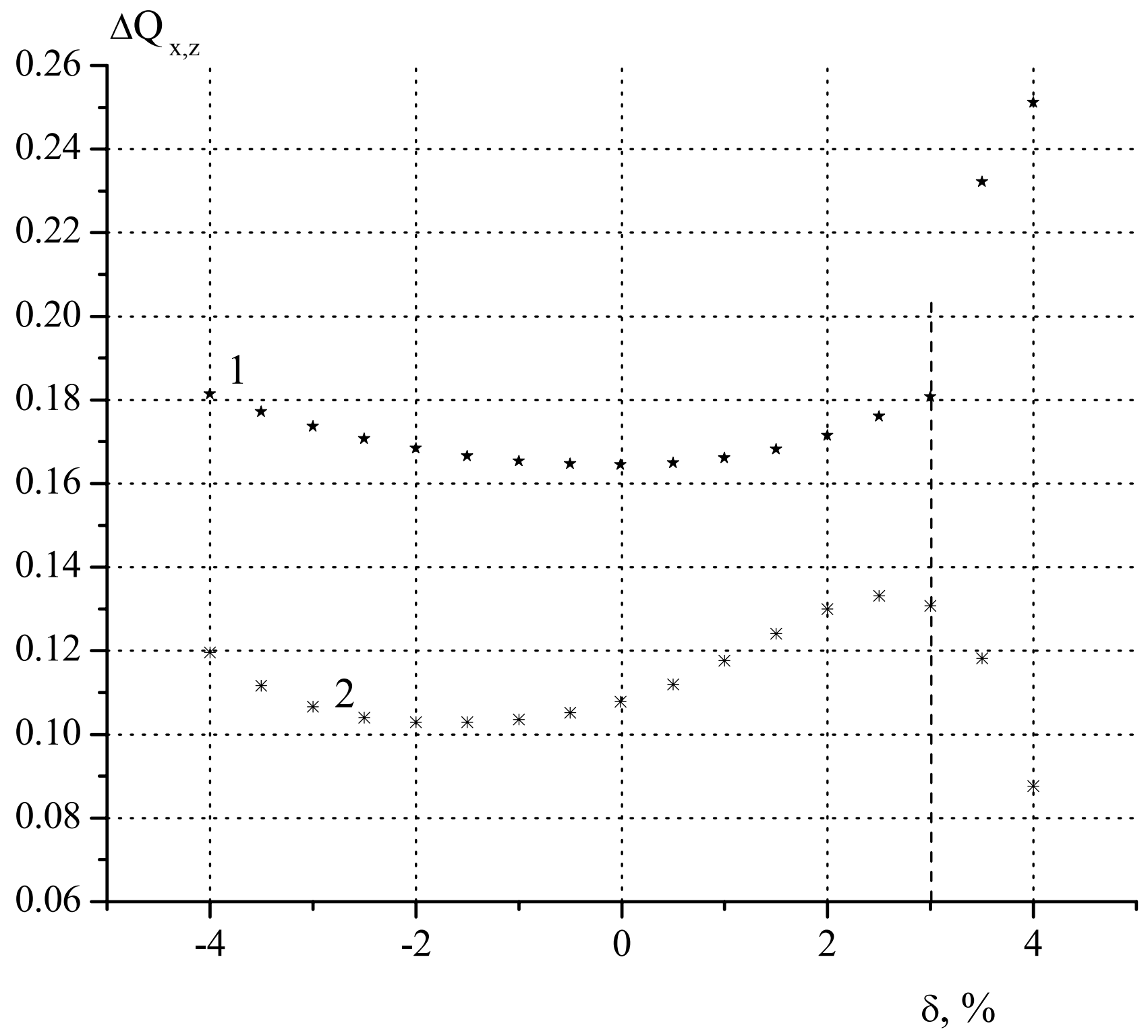

Fig.11. Fractional part of betatron frequencies vs. momentum deviation. Betatron amplitudes at IP are $x_b = 1$ mm, $z_b = 0.5$ mm. Vertical dash line bounds range of stable beam motion for $\delta > 0$.

Biological and medical studies will be ones of main application of LESR. X-rays with photon energy about of 33 keV are needed for coronary angiography and electron beam energy about of 43 MeV is needed to obtain such photons under head-on collision. X-rays with photon energies over range 6 keV $< \varepsilon_\gamma <$16 keV and with long-term stability of the intensity are needed for many biological researches. We need to use electron beam with energies over range 18 MeV $< E_0 <$ 30 MeV in order to generate such X-rays. The stable storage ring operation with intensive electron beam is practically impossible at such small energies because of IBS. The dependences of the horizontal and longitudinal emittances on time for stored bunch current $I_{stor} = 10$ mA and electron beam energy $E_0 = 43$ MeV are presented in Fig.12. The parameters of injected beam are following: bunch charge $q_b = 0.5$ nC, emittances $\varepsilon_x = \varepsilon_z = 10^{-7}$, $\varepsilon_s = 6*10^{-6}$. During 0.1 s between injection pulses horizontal and longitudinal emittance grow about of one order and three times, accordingly, what causes the essential decreasing of the X-rays intensity. The steady-state emittances are $\varepsilon_x \approx 5*10^{-6}$, $\varepsilon_s \approx 1.5*10^{-5}$ at coupling coefficient $\kappa = 0.05$. Under such conditions X-rays intensity is approximately $10^{11}$ phot /s and such intensity is not acceptable in angiographical studies. The problems drastically complicate under electron beam energy decreasing.

To meet the requirements of the medical and biological studies we intend to use the following operation modes of the storage ring. Two optical cavities with the length $L_{res} = \lambda_{RF} \approx 430$ mm will be placed inside of the final lenses of the quadruplet as it is shown in Fig.13. The crossing angles are equal to $\varphi_1 = 10°$ and $\varphi_2 = 150°$. Under such lengths of the optical cavities we will be able to use 18 electron bunches of the storage ring. The medical photons will be generated at electron beam energy $E_0 \approx 43$ MeV and collision angle $\varphi_1 = 10°$. Electron beam with non-steady-state size will be used in this mode. In order to achieve the high average X-rays intensity ($n_\gamma \geq 10^{12}$ / s) electron beam will be injected in storage ring with repetition frequency $f_{inj} = 10$ Hz. The biological photons with energy over range 6 keV $< \varepsilon_\gamma <$16 keV will be generated at electron beam energy 70 MeV < $E_0$ < 120 MeV and collision angle $\varphi_2 = 150°$. Of course, X-rays intensity at such collision angle will be much less than the one at small collision angle. Nevertheless, this intensity quite meets the requirements of the biological experiments. Besides, X-rays intensity will be very stable because we intend to use electron beam with steady-state parameters (at electron beam energy about of 100 MeV IBS-effects appear low). In both operation modes we are going to use the neodymium laser with photon energy $\varepsilon_{las} = 1.164$ eV.

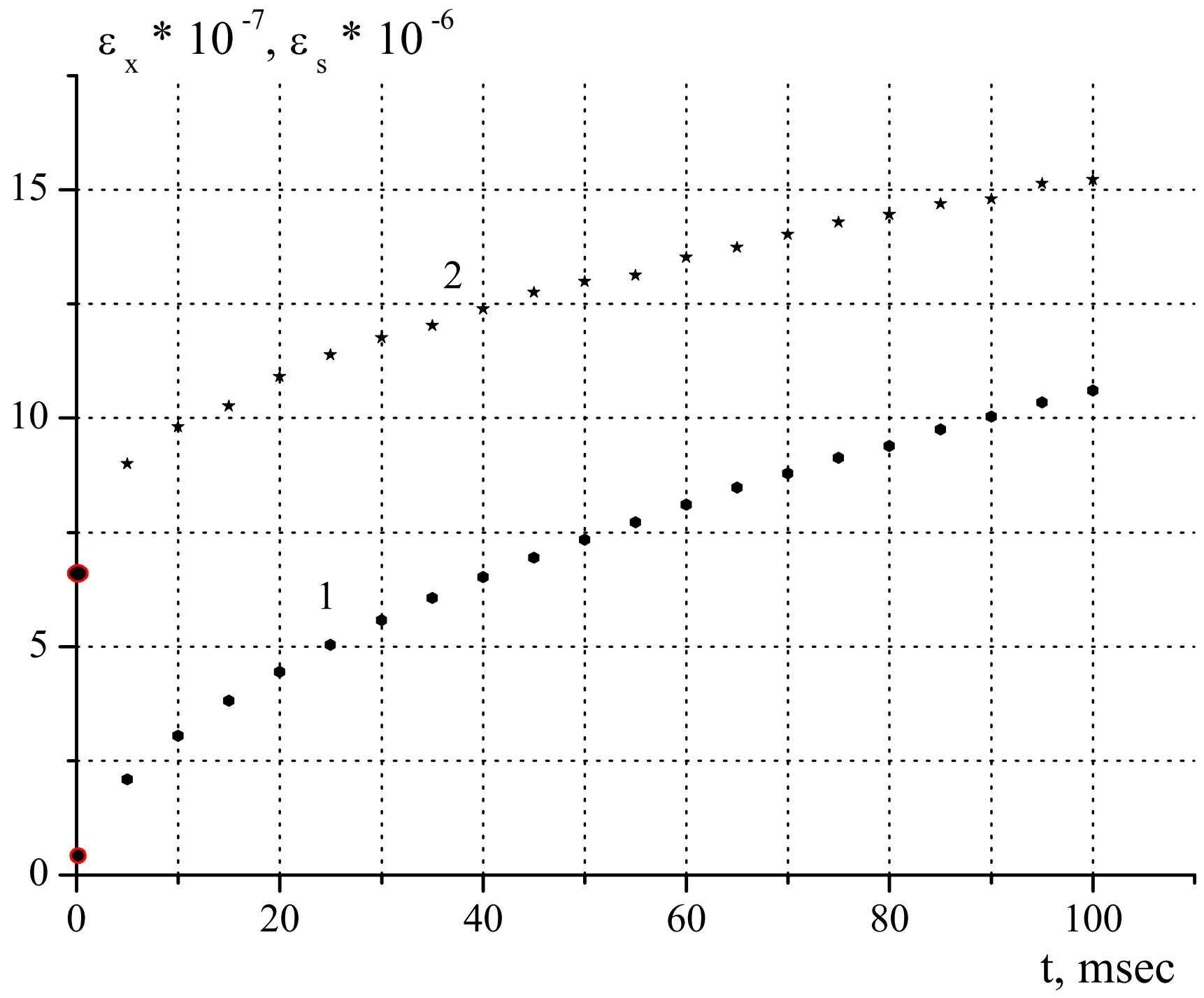

Fig.12. Horizontal (1) and longitudinal (2) emittances during injection interval. Electron beam energy $E_0 = 43$ MeV, coupling coefficient $\kappa = 0.05$, bunch current $I_{stor} = 10$ mA.

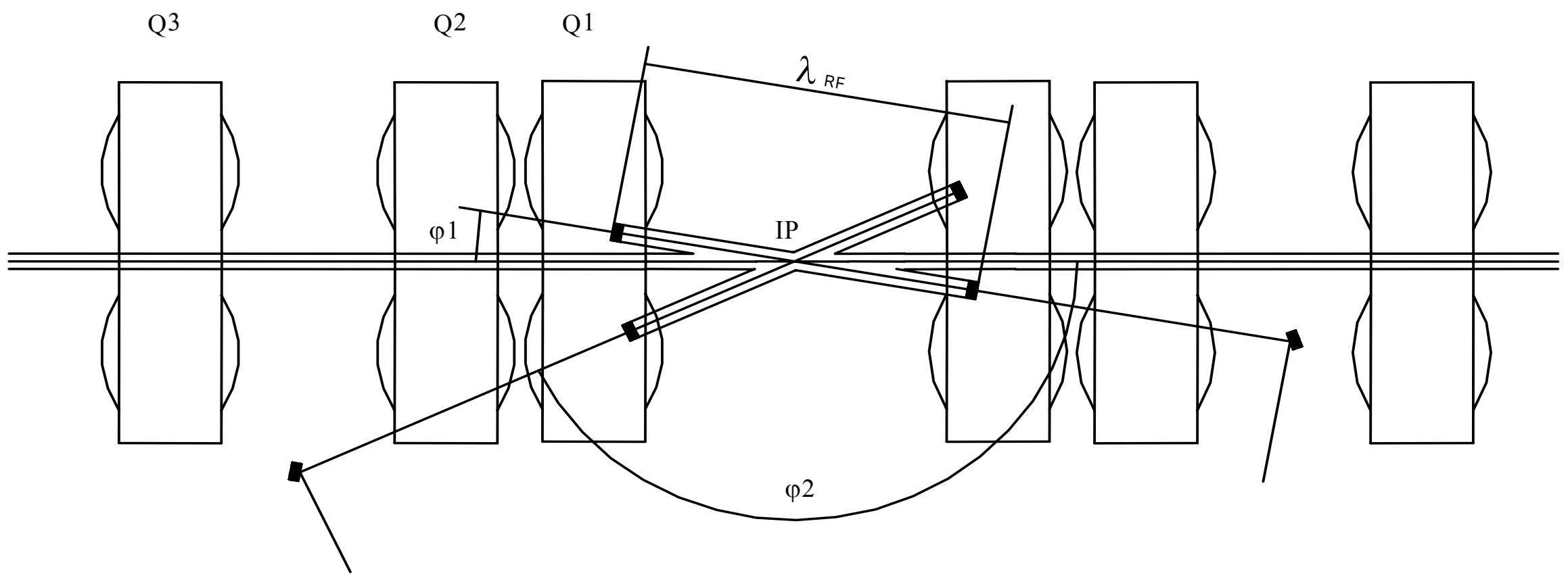

Fig.13. Arrangement of the equipment on IP drift. Collision angles $\varphi_1 = 10°$ and $\varphi_2 = 150°$.

Several following figures illustrate all above stated (all dependencies were calculated by electron beam dynamics simulation in chosen lattice involving Compton scattering). The computed intensity of the medical photons during injection interval is shown in Fig.14, spectrum of the scattered photons within $4\pi$-solid angles is shown in Fig.15. The average intensities during injection interval and within 10 ms period just after injection are approximately equal to $1.2 \cdot 10^{12}$ phot /s and $2 \cdot 10^{12}$ phot /s, accordingly. The spectral brightness under laser beam size $\sigma_{las} = 40\ \mu$ is approximately equal to $5*10^{12}$ phot / (s*mrad*mm$^2$*0.1%BW) and such brightness allows to carry out angiographic studies.

The collimated spectrum of the photons for biological studies with maximal energy $\varepsilon_{\gamma max} \approx 6.7$ keV (electron beam energy $E_0 = 75$ MeV, collision angle $\varphi = 150°$) is presented in Fig.16. Total Compton beam intensity within $4\pi$-solid angle is approximately equal to $n_\gamma \approx 10^{11}$ / s, number of photons within collimation angle $\alpha_{col} = 1$ mrad is $n_{\gamma col} \approx 2.5*10^9$ / s, spectrum width (FWHM) is approximately 10%, number of photons within 0.1%BW at maximal photon energy is $n_{\gamma BW} \approx 2.3*10^8$ / s. Under laser beam waist $\sigma_{las} = 40\ \mu$ the spectral brightness is $B \approx 2*10^{11}$ / (s*mm$^2$*mrad*0.1 % BW). X-rays with such parameters quite meet the requirements of biological studies.

In order to generate hard $\gamma$-quanta we intend to use 10°-collision of laser photons and high-energy electron beam with steady-state parameters. The total Compton beam intensity at maximal operation energy of the storage ring $E_{0max} = 225$ MeV is shown in Fig.17. Maximal $\gamma$-quanta energy is $\varepsilon_{\gamma max} \approx 900$ keV ($\varepsilon_{las} = 1.164$ eV). Insignificant decreasing of the scattered beam intensity is caused by energy spread increasing and electron bunches lengthening under Compton scattering. By means of the "green laser" ($\varepsilon_{las} = 2.328$ eV) we will be able to obtain 1.8 MeV $\gamma$-quanta energy. Such $\gamma$-quanta, for example, may be used for neutron generation in beryllium target.

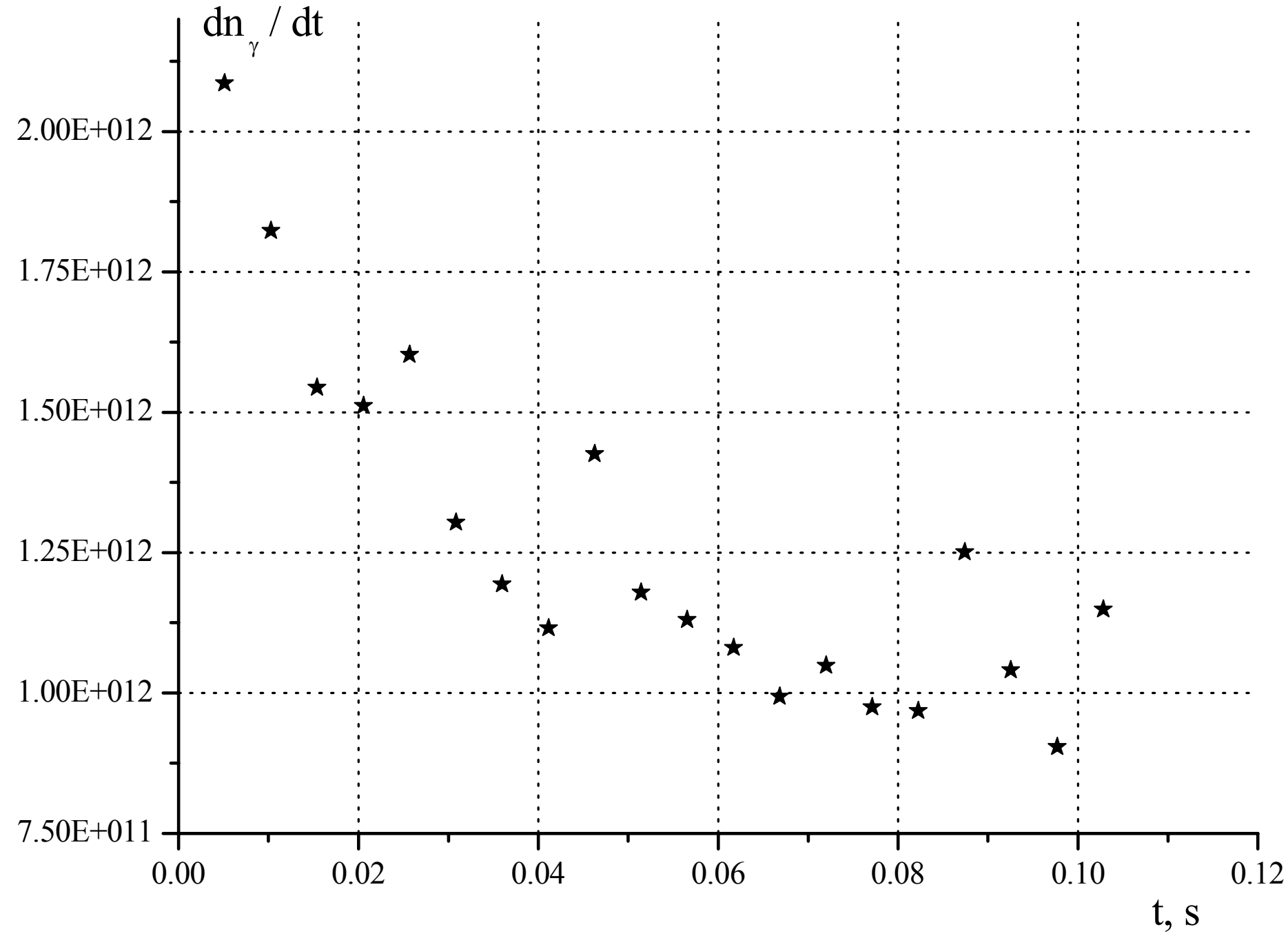

Fig.14. Compton beam intensity during injection interval. Electron beam energy $E_0 = 43$ MeV, bunch stored current $I_{stor} = 10$ mA, bunch number $n_b = 18$, stacked laser flash energy $w_{las} = 1$ mJ, collision angle $\varphi = 10°$, laser beam size $\sigma_{las} = 40\ \mu$.

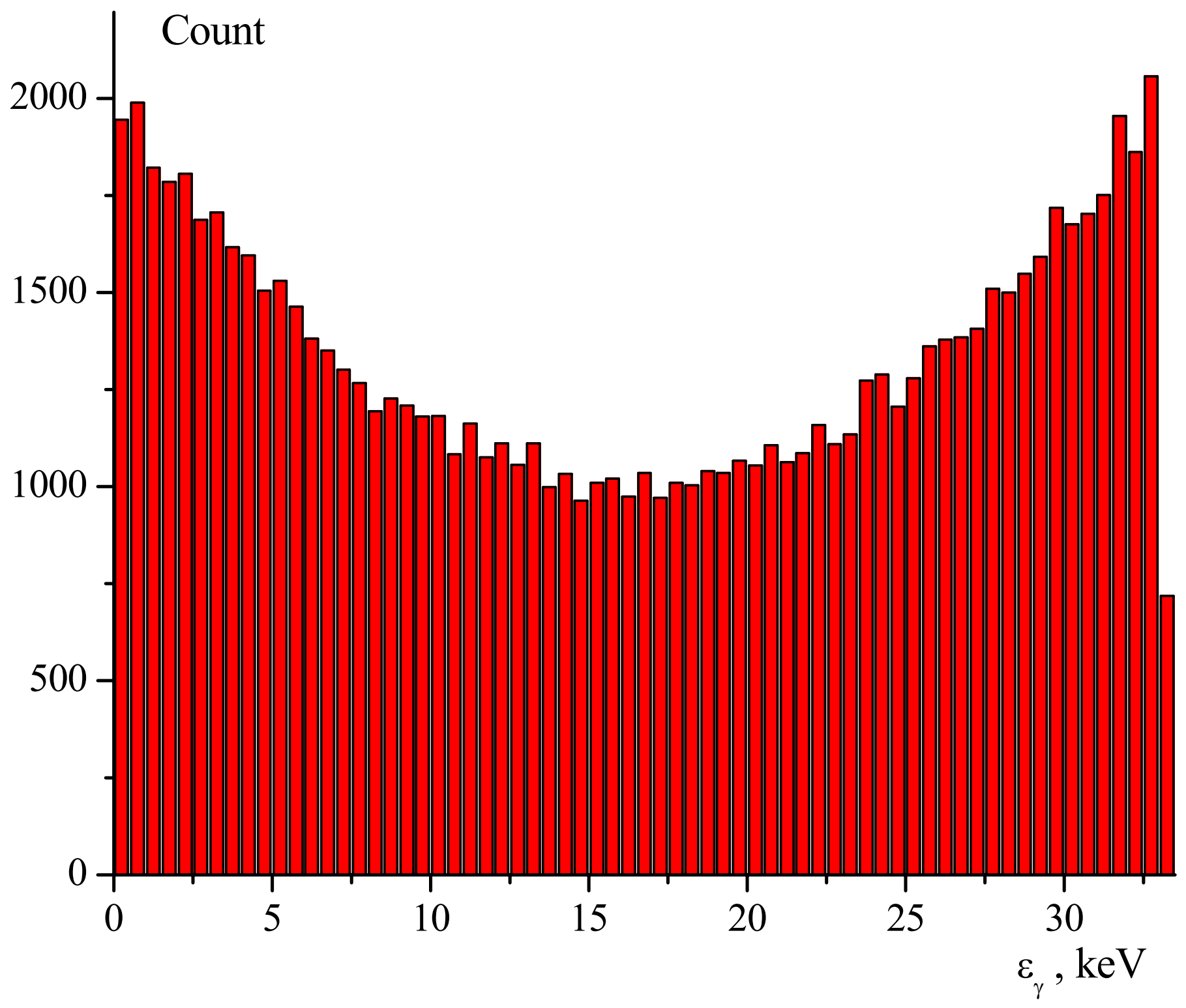

Fig.15. Spectrum of scattered photons within 4π-solid angle. Electron beam energy $E_0$ = 43 MeV, collision angle φ = 10°.

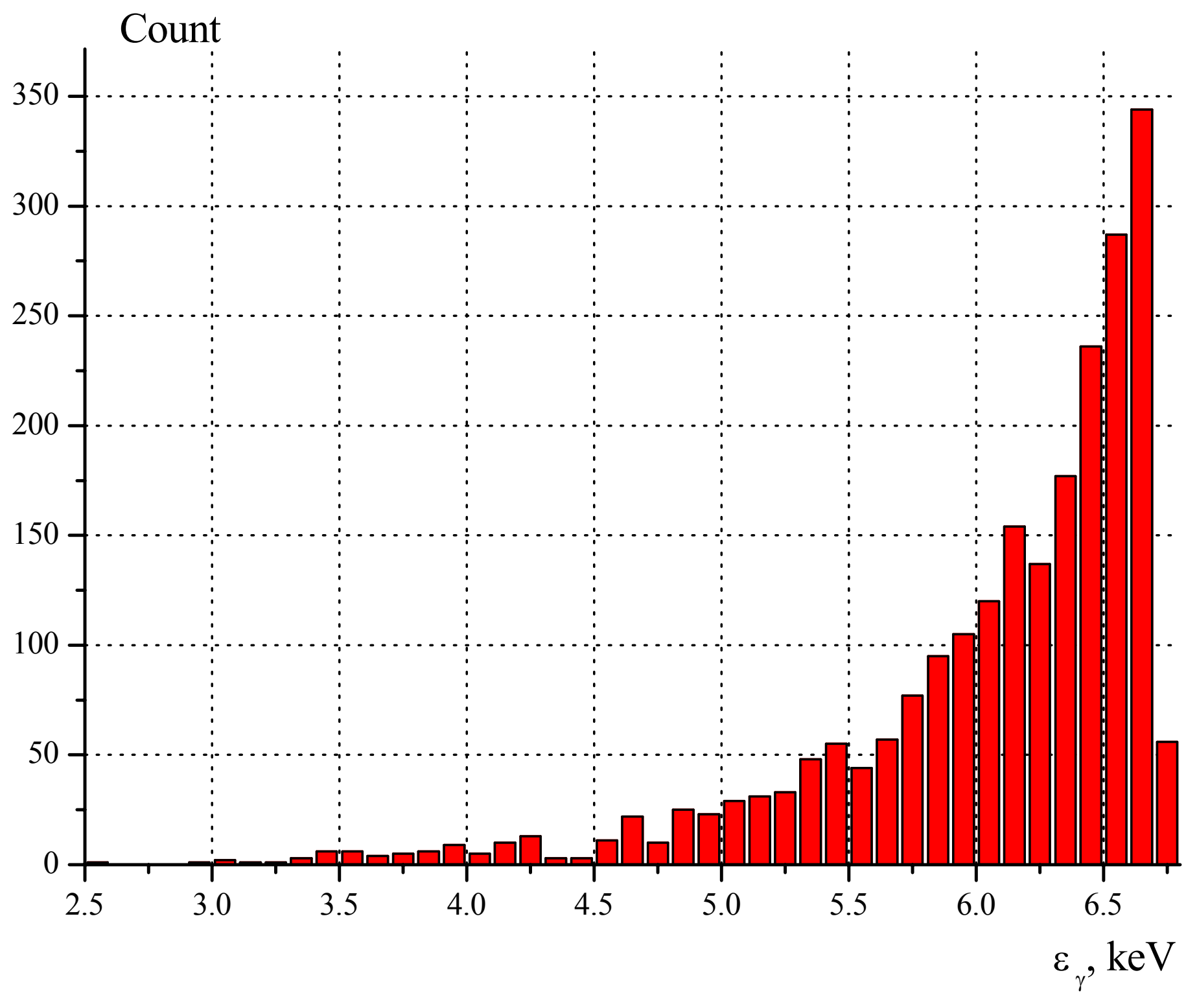

Fig.16. Scattered photons spectrum within collimation angle $\alpha_{col}$ = 1 mrad. Electron beam energy $E_0$ = 75 MeV, collision angle φ = 150°.

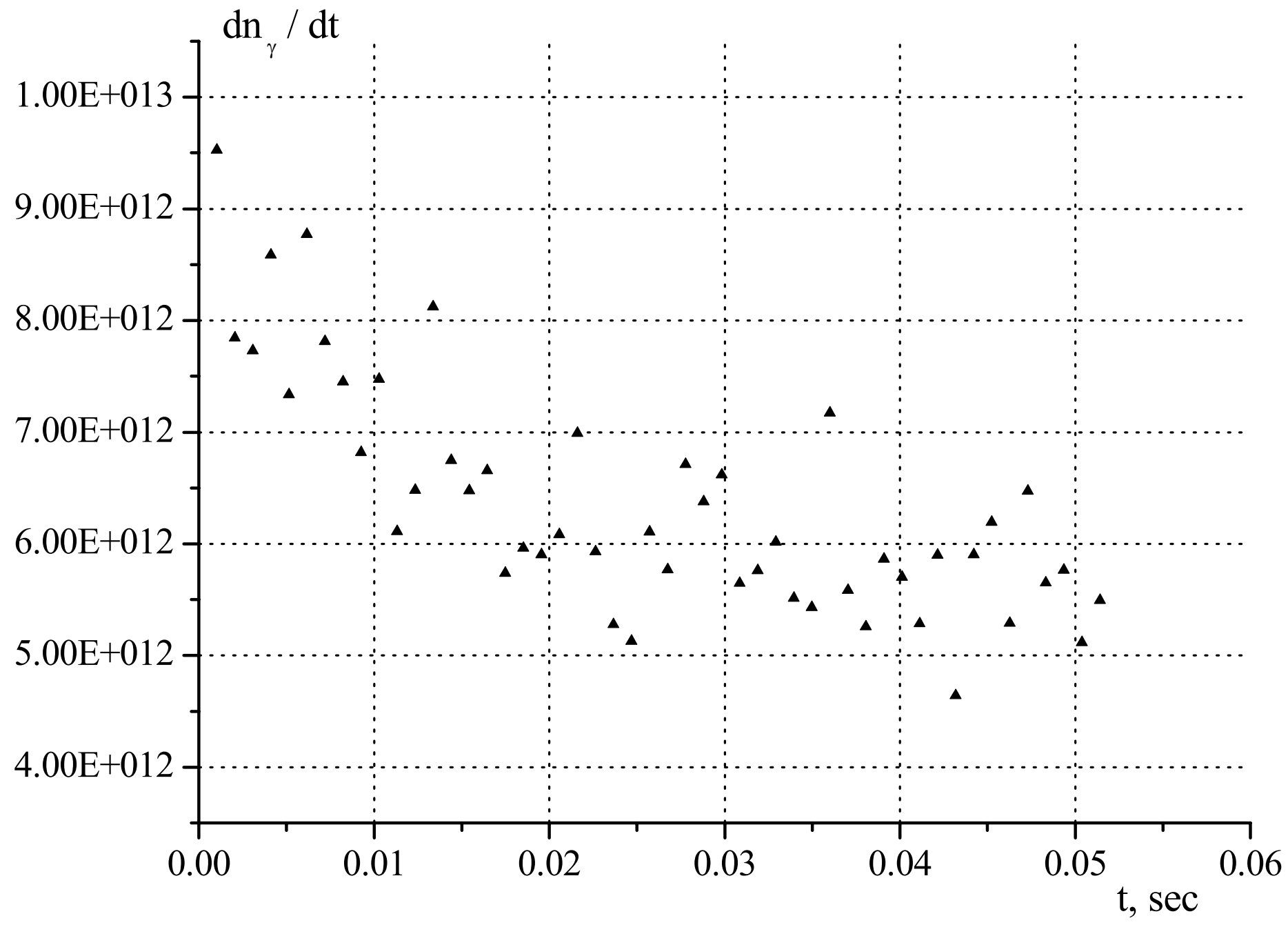

Fig.17. Compton beam intensity. Electron beam energy $E_0$ = 225 MeV, stored bunch current $I_{stor}$ = 10 mA, number of electron bunches $n_b$ = 18, stacked laser flash energy $w_{las}$ = 1 mJ, collision angle $\varphi = 10°$.

The dynamics aperture of the storage ring at IP azimuth for linear momentum compaction factors $\alpha_1 = 0.01$ and $\alpha_1 = 0.02$ is presented in Fig.18.

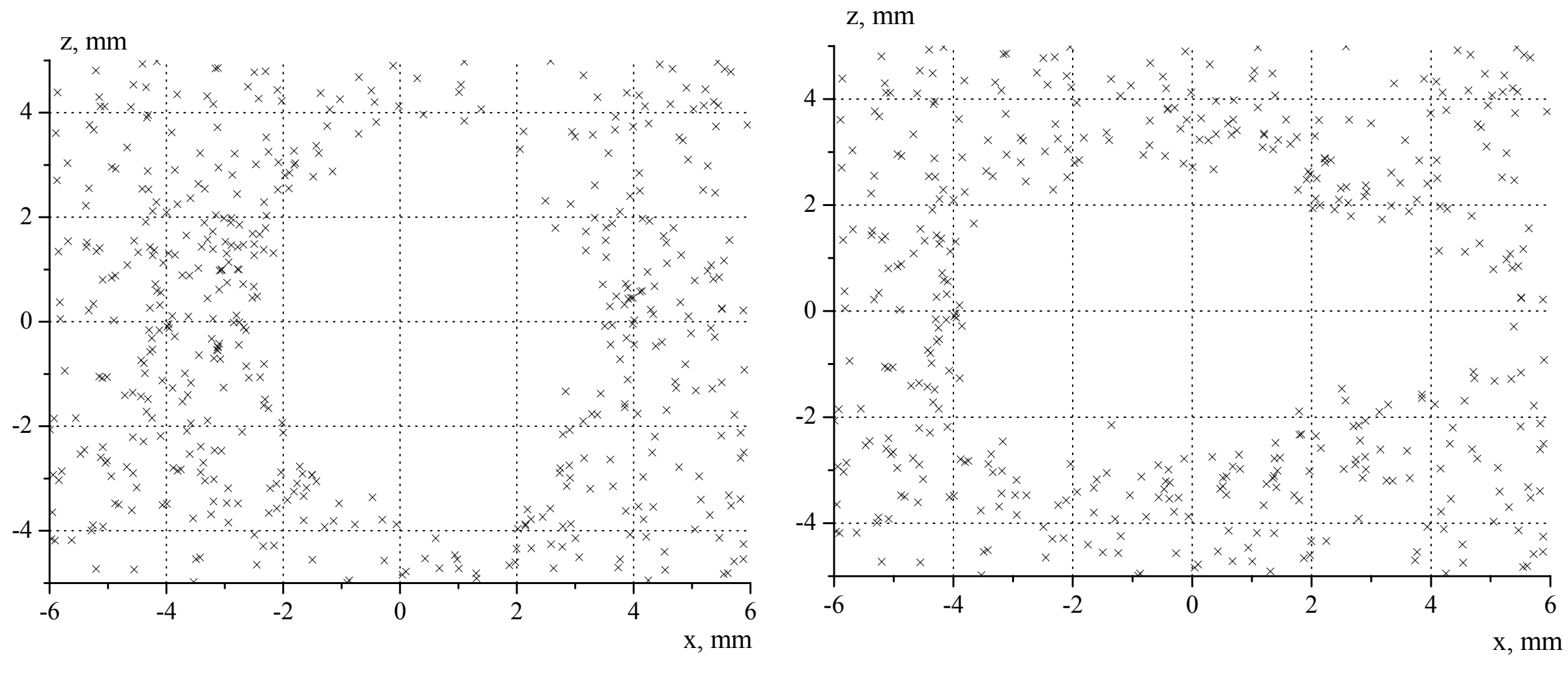

a b

Fig.18. Dynamics aperture of storage ring for operation modes with momentum compaction factors $\alpha_1 = 0.01$ (a) and $\alpha_1 = 0.02$ (b). Particles momentum deviation is $\delta = 3$ %

Together with large energy acceptance such DA will allow the storing of the intensive electron beam and, consequently, the obtaining of the intensive X-rays.

## *5. Summary*

The problems of the electron beam dynamics associated with large beam energy spread caused by intensive Compton scattering are solved in proposed lattice of the laser-electron storage ring. X-rays over energy range 6 keV $\leq \varepsilon_\gamma \leq$ 900 keV with long-term stable intensity up to $10^{13}$ phot /s may be generated under realized parameters of the injector, storage ring and laser system.

Maximum allowed Compton beam intensity limited by energy acceptance of the storage ring is approximately $10^{15}$ phot /s over all energy range.

The main storage ring, electron and Compton beam parameters are presented in Tabl.1.

Table 1. The main ring, electron and Compton beam parameters

| Parameter | Value |
|---|---|
| Circumference, m | 15.418 |
| Energy range, MeV | 40-225 |
| Betatron tunes $Q_x$ , $Q_z$ | 3.155; 2.082 |
| Amplitude functions $\beta_x$, $\beta_z$ at IP, m | 0.14; 0.12 |
| Linear momentum compaction factor $\alpha_1$ | 0.01-0.078 |
| RF acceptance, % | > 5 |
| RF frequency, MHz | 700 |
| RF voltage, MV | 0.3 |
| Harmonics number | 36 |
| Number of circulating electron bunches | 2; 3; 4; 6; 9; 12; 18; 36 |
| Electron bunch current, mA | 10 |
| Stacked laser flash energy into optical cavity, mJ | 1 |
| Collision angle, degrees | 10; 150 |
| Scattered photon energy (Nd laser, $\varepsilon_{las}$ = 1.16 eV), keV | 6-900 |
| Maximum scattered photon intensity, phot /s | up to $10^{13}$ |